\documentstyle[11pt,newpasp,twoside,epsf]{article}
\begin{document}
\title{Low Frequency Insights Into Supernova Remnants}
 \author{Kristy K. Dyer, Stephen P. Reynolds, Kazik J. Borkowski}
\affil{North Carolina State University, Physics Department Box 8202, Raleigh NC 27695}
\author{Namir E. Kassim, Christina K. Lacey}
\affil{Naval Research Lab, Remote Sensing Division, Code 7213 Washington DC 20375-5351}

\begin{abstract}
Low frequency observations at 330 and 74 MHz can provide new insights into
supernova remnants (SNR). We can test theoretical predictions for spectral
index variations. Nonlinear models of shock acceleration predict that the
spectra from young SNR should be slightly concave rather than power laws
-- flattening toward higher energies. However, few SNR are bright and
compact enough to be studied at millimeter wavelengths, restricting
studies to the small range from 6 to 20 cm (a factor of 1.7 in electron
energies). Observations at 330 MHz increase the electron energy baseline
to a factor of 4, while providing sensitivity to larger spatial scales
that are resolved out by centimeter-wavelength interferometers.  Such
observations can also separate thermal from nonthermal emission and detect
excess free-free absorption associated with cool gas in remnants. Wide
field images also provide an efficient census of both thermal and
nonthermal sources over a large region.
\end{abstract}

\section{Introduction}

In order to understand shocks in supernova remnants (SNR) we need to separate three issues: intrinsic properties of the explosions themselves, the character of the SNR environment, and observational constraints. In order to obtain fundamental facts about the explosion such as the age and energy released we must understand the structure of the circumstellar medium. In order to interpret observations we must understand observational limits imposed by the nature of single dish and interferometric observations.

Low frequency interferometric observations can help disentangle these three overlapping issues and will have the opportunity to contribute to three multifrequency issues: finding X-ray synchrotron emission, measuring spectral curvature predicted by particle theory, and clearing up uncertainties in observed spectral index variations.

\section{Information from Total Flux Measurements}

In order to understand supernovae (SN) and SNR, we must separate SNR from their environs. Since SNR are found preferentially near star forming regions, in the galactic plane, they are, not by chance, often in complex regions of the sky (for a convincing example consult the galactic plane surveys by Effelsberg: Reich, Reich, \& Fuerst, 1990, 1997). 

Even something as simple as measuring the total flux from a SNR requires imaging to avoid confusing the emission from nearby sources. While interferometers can over resolve the remnant, losing total flux information, fluxes obtained from single dish measurements often confuse the SNR with nearby objects.

An individual electron of energy E radiates its peak synchrotron emission at frequency $\nu \propto E{^2} B$. Since frequency is proportional to the energy squared, we need the leverage of much wider frequency ``baselines'' to study subtle changes in the electron spectrum. Comparing observations from 6 to 20 cm is only a range of 1.7 in energy, whereas 74 MHz to 4.6 GHz buys a factor in energy of 8.

Low frequencies also avoid a scale problem. For the VLA, many galactic SNR are large enough they are over-resolved at wavelengths longer than 6 cm. Total flux is more reliably measured at at lower frequencies.  While low-frequency interferometric observations can be affected by absorption, even a lower limit to the flux would help firm up the predictions.

\section{X-ray Synchrotron Emission}

X-ray emission in SNR is generally considered to be thermal; however, certain SNR look suspiciously similar in the X-ray and radio (such as G41.1-0.2; see Dyer \& Reynolds 1999). Morphological similarity does not prove the X-rays are synchrotron -- but at the very least it suggests that X-rays are being excited at the same location as the relativistic electrons that produce radio synchrotron emission.

In fact, X-ray observations of some SNR like SN1006 (Koyama et al 1993) and  (Slane et al. 2000) show the spectra are dominated by synchrotron emission. A more serious threat to our understanding of shocks is the possibility that other SNR could have a smaller synchrotron component confusing the thermal emission -- this would stymie thermal fits and prevent accurate measurements of shock temperatures and elemental abundances.

Models have been developed by Reynolds (1993, 1996) to describe this emission. Two simple models, {\it SRCUT} and {\it SRESC}, are available in XSPEC 11.0. These models rely on the radio flux and spectral index as reported by Green (1998).  The models differ subtly -- the precise shape of the X-ray synchrotron spectrum can be used to determine properties of the SNR, including the age of the shock, magnetic fields and electron energies and synchrotron losses. The models depend on accurate extrapolations of the radio synchrotron spectrum over eight orders of magnitude of frequency.
The current state of this knowledge is very poor, as can be seen from examining collections of flux measurements (Truskin 1999, see G041.1-0.2 for example). Reported fluxes can vary by a factor of 1/3 to 2, sometimes even between measurements made by the same instrument. Most single dish instruments do not have the resolution to separate SNR from nearby sources and absolute fluxes are not well calibrated from one instrument to another. The uncertainties reported in the literature are often absent or optimistic underestimates. Low frequency interferometric observations can contribute reliable measurements with (most importantly) accurate uncertainties, allowing us to separate thermal and non-thermal X-rays.

\section{Spectral Curvature}

Non-linear first-order Fermi shock acceleration has been shown to be the leading model describing particle acceleration in SNR shocks. Since protons determine the shock structure, in the past particle codes studying shocks have ignored electrons. However it is commonly held that higher energy electrons have longer scattering lengths across the shock. If this is true, electrons interact differently with the shock. Highly energetic electrons see a shock with a higher compression ratio than low energy electrons, and therefore gain more energy than their low energy counterparts. Tests with particle codes including electrons by Ellison \& Reynolds (1991, 1992) showed the synchrotron spectrum, while very close, is not exactly a powerlaw -- it deviates very slightly -- concave upwards or flattening to higher energies. This subtle curvature had already been found observationally in single dish measurement of well studied remnants such as Tycho and Kepler.

This is one of the few methods by which limits can be set on the magnetic field {\bf independent} of the electron energy, putting  us closer to to the goal of finding intrinsic properties of the SNR. In some cases a single accurate measurements at low frequency can discriminate between models with different magnetic fields. 

\section{Spectral Index Variations and Inherent Problems} 

It is worth noting that the spectral index variations theorists look for, to obtain insight on shock mechanisms, should be very small. SNR look very similar from one frequency to the next. In addition, if parts of the remnant varied widely it would be unlikely that the spatial average would come as close as it does to a power law over three orders of magnitude in frequency. 


Studies of spectral index variations across the face of the remnant bring out the worst in interferometric measurements. There are two serious problems underlying spectral index fluctuations reported in the literature. First, even with scaled arrays, interferometric observations at different frequencies have slightly different UV coverage. This difference is compounded by processing with non-linear deconvolution methods. Second, if we are to believe the small effects we are looking for, we must be able quantify the noise accurately -- and the noise on extended sources, processed through CLEAN or MEM, is not well understood. A 3$\sigma$ effect is meaningful only if $\sigma$ is well known. 
We have found that re-observing a SNR with the VLA with slightly better UV coverage found spectral index variations on the same scale as previous observations: however these variations were in different locations with different signs (Dyer \& Reynolds 1999). This was true even when linear regression algorithms were used, designed to take into account an offset due to lack of short-spacing information. 

However some SNR do show statically significant variations (\& Green 19??). The situation can be improved by adding single dish data, deriving indices from observations at three or four frequencies rather than two (including lower frequencies), and by testing algorithms designed to avoid the zero spacing problem and finally by better understanding of the statistical noise across diffuse CLEANed emission. 

The last two issues could be addressed by a small but critical project -- The techniques used to find spectral index variations (regression methods, T-T plots and spectral tomography) in SNR could be used to look for spectral index variations where we know there should be none -- in thermal H II regions such as the Orion nebula. A thermal nebula should have a flat spectrum ($\nu ^{-0.1}$) with no spectral index variations beyond statistical fluctuations, therefore the variations found would tell us something about the noise in our spectral index maps of SNR. A thermal nebula also provides extended emission where the unknown effects of CLEAN on source noise could be checked.

\end{document}